\documentclass[]{aastex631}

\pdfoutput=1 \usepackage{amsmath,amstext}
\usepackage[T1]{fontenc}
\usepackage{apjfonts} 
\usepackage[figure,figure*]{hypcap}
\usepackage{hyperref}
  \newcommand{\p}[1]{{\color{cyan}{#1}}}
\renewcommand{\d}[1]{{\color{red}{#1}}}

\newcommand{\angstrom}{\mbox{\normalfont\AA}}
\newcommand{\g}[1]{{\color{cyan}{#1}}}

\shorttitle{Subchromospheric Reconnection}
\shortauthors{Baker, D. et al.}

\begin{document}

\title{Searching for evidence of subchromospheric magnetic reconnection on the Sun}

\author[0000-0002-0665-2355]{Baker, D.}
\affiliation{University College London, Mullard Space Science Laboratory, Holmbury St. Mary, Dorking, Surrey, RH5 6NT, UK}
\author{van Driel-Gesztelyi, L.}
\affiliation{University College London, Mullard Space Science Laboratory, Holmbury St. Mary, Dorking, Surrey, RH5 6NT, UK}
\affiliation{LESIA, Observatoire de Paris, Universit\'e PSL, CNRS, Sorbonne Universit\'e, Univ. Paris Diderot, Sorbonne Paris Cit\'e, 5 place Jules Janssen, 92195 Meudon, France}
\affiliation{Konkoly Observatory, Research Centre for Astronomy and Earth Sciences, Konkoly Thege \'ut 15-17., H-1121, Budapest, Hungary}
\author[0000-0001-7927-9291]{James, A. W.}
\affiliation{University College London, Mullard Space Science Laboratory, Holmbury St. Mary, Dorking, Surrey, RH5 6NT, UK}
\author{D\'emoulin, P.}
\affiliation{LESIA, Observatoire de Paris, Universit\'e PSL, CNRS, Sorbonne Universit\'e, Univ. Paris Diderot, Sorbonne Paris Cit\'e, 5 place Jules Janssen, 92195 Meudon, France}
\affiliation{Laboratoire Cogitamus, rue Descartes, 75005 Paris, France}
\author{To, A. S. H.}
\affiliation{ESTEC, European Space Agency, Keplerlaan 1, PO Box 299, NL-2200 AG Noordwijk, The Netherlands}
\affiliation{University College London, Mullard Space Science Laboratory, Holmbury St. Mary, Dorking, Surrey, RH5 6NT, UK}
\author{Murabito, M.}
\affiliation{INAF Istituto Nazionale di Astrofisica, Osservatorio Astronomico di Roma, 00078, Monte Porzio Catone (RM), Italy
}
\affiliation{Space Science Data Center (SSDC), Agenzia Spaziale Italiana, via del Politecnico, s.n.c., I-00133, Roma, Italy
}
\author{Long, D.}
\affiliation{School of Physical Sciences, Dublin City University, Glasnevin Campus, Dublin, D09 V209, Ireland}
\author{Brooks, D. H.}
\affiliation{Department of Physics $\&$ Astronomy, George Mason University, 4400 University Drive, Fairfax, VA 22030, USA}
\affiliation{University College London, Mullard Space Science Laboratory, Holmbury St. Mary, Dorking, Surrey, RH5 6NT, UK}
\author[0000-0002-4071-5727]{McKevitt, J.}
\affiliation{University College London, Mullard Space Science Laboratory, Holmbury St. Mary, Dorking, Surrey, RH5 6NT, UK}
\affiliation{University of Vienna, Institute of Astrophysics,
Türkenschanzstrasse 17,
Vienna 1180, Austria}
\author[0000-0002-3362-7040]{Laming, J. M.}
\affiliation{Space Science Division, Naval Research Laboratory, Code 7684, Washington, DC 20375, USA}
\author{Green, L. M.}
\affiliation{University College London, Mullard Space Science Laboratory, Holmbury St. Mary, Dorking, Surrey, RH5 6NT, UK}
\author[0000-0003-2802-4381]{Yardley, S. L.}
\affiliation{Department of Mathematics, Physics and Electrical Engineering, Northumbria University, Newcastle Upon Tyne, NE1 8ST, UK}
\affiliation{Department of Meteorology, University of Reading, Reading, UK}
\affiliation{Donostia International Physics Center (DIPC), Paseo Manuel de Lardizabal 4, San Sebasti{\'a}n, 20018, Spain}
\author{Valori, G.}
\affiliation{Max Planck Institute for Solar System Research, Justus-von-Liebig-Weg 3, 37077 G\"ottingen, Germany}

\author{Mihailescu, T.}
\affiliation{University College London, Mullard Space Science Laboratory, Holmbury St. Mary, Dorking, Surrey, RH5 6NT, UK}
\author{Matthews, S. A.}
\affiliation{University College London, Mullard Space Science Laboratory, Holmbury St. Mary, Dorking, Surrey, RH5 6NT, UK}
\author{Kuniyoshi, H.}
\affiliation{Department of Earth and Planetary Science, The University of Tokyo, 7-3-1 Hongo, Bunkyo-ku, Tokyo 113-0033, Japan}

\begin{abstract}
Within the coronae of stars, abundances of those elements with low first ionization potential (FIP) often differ from their photospheric values. 
The coronae of the Sun and solar-type stars mostly show enhancements of low-FIP elements (the FIP effect) while more active stars such as M dwarfs have coronae generally characterized by the inverse-FIP (I-FIP) effect. 
Highly localized regions of I-FIP effect solar plasma have been observed by Hinode/EIS in a number of highly complex active regions, usually around strong light bridges of the umbrae of coalescing/merging sunspots.
These observations can be interpreted in the context of the ponderomotive force fractionation model which predicts that plasma with I-FIP effect composition is created by the refraction of waves coming from below the plasma fractionation region in the chromosphere. 
A plausible source of these waves is thought to be reconnection in the (high-plasma $\beta$) subchromospheric magnetic field.
In this study, we use the 3D visualization technique of \cite{chintzoglou13} combined with observations of localized I-FIP effect in the corona of AR 11504 to identify potential sites of such reconnection and its possible consequences in the solar atmosphere.
We found subtle signatures of episodic heating and reconnection outflows in the expected places, in between magnetic flux tubes forming a light bridge, within the photosphere of the active region.
Furthermore, on either side of the light bridge, we observed small antiparallel horizontal magnetic field components supporting the possibility of reconnection occuring where we observe I-FIP plasma. 
When taken together with the I-FIP effect observations, these subtle signatures provide a compelling case for indirect observational evidence of reconnection below the fractionation layer of the chromosphere, however, direct evidence remains elusive.
\end{abstract}


\section{Introduction}\label{s:intro}

The elemental composition of the Sun's corona, like other solar-type stars, is dominated by the First Ionisation Potential (FIP) effect, i.e. in most coronal structures the abundances of elements with FIP $\leq$10 eV are enhanced relative to their photospheric abundances. 
Although there is a large  variation \citep{mihailescu22}, active regions (ARs) at or just after their magnetic flux peak show the highest enhancements  (FIP bias) whereas open-field coronal holes show FIP bias close to 1, i.e. they exhibit similar chemical composition in the photoshpere and in the corona.

It came as a great surprise that inverse-FIP (I-FIP) composition was detected in the solar corona in patches over (or in the close vicinity of) sunspot umbrae in the late phase of flares \citep{doschek15,doschek16,doschek17,baker19,baker20}. 
I-FIP means depleted/increased abundance of low-FIP/high-FIP elements, which is common in the coronae of active cool stars like {\bf M} dwarfs \citep[e.g.][and references therein]{laming15}. 

In the solar atmosphere, it was shown that the I-FIP effect is due to depletion of low-FIP elements \citep{brooks18}, in agreement with the ponderomotive force  
fractionation model.
The model explains the generation of I-FIP plasma by fast-mode waves propagating upward from the photosphere and refracting/reflecting at the large density gradient in the chromosphere, where low-FIP elements are mostly ionised 
\citep{laming15,laming21}. \cite{baker19,baker20} analysed the magnetic evolution of ARs in which I-FIP composition was observed and found that most of the I-FIP patches are located above forming (merging) umbrae.
This plasma is then evaporated into the corona when the flare ribbons cross the umbrae.
They proposed that such merging umbrae are likely to undergo subsurface magnetic reconnection as the emerging magnetic threads are being pushed together by buoyancy. 
The acoustic waves created during the reconnection process will mode convert at the plasma $\beta$=1 layer to fast-mode waves \citep[e.g.][]{schunker06,cally07} that will deplete low-FIP ions from the chromosphere at their reflection location \citep{laming21}. 

Magnetic reconnection in the corona - where the majority of our observations originate - is taking place in a  low-$\beta$ environment. Although a direct observation of reconnection remains elusive, there is plenty of indirect evidence from multi-wavelength observations as well as from theory and simulations that it is taking place in the solar corona and solar wind and it drives most activity phenomena we observe there \citep[e.g.][]{benz17,pontin22,russel23}. 

The subsurface magnetic reconnection invoked by \cite{baker19,baker20} is assumed to take place below the optical depth $\tau =1$ layer, therefore it has hardly any observational evidence, not even indirect, besides the I-FIP coronal composition above merging umbrae \citep{pontin22}. Analytical computations \citep{parker78,parker79} and simulations clearly show, however, that magnetic reconnection is perfectly possible between flux tubes/ropes below the photosphere \citep[e.g.][]{linton06,murray07,tortosa09,prior16,toriumi17,toriumi19}.
The plasma $\beta$ changes rapidly with depth and can reach $\beta\approx 10^4$ at 2 Mm depth \citep{chen22}. 
However, the initiation and effects of reconnection differ from the low-$\beta$ case in that the field lines must be brought together by plasma motions. 
The main differences in effects of energy release as a function of plasma $\beta$ were investigated by \cite{peter19} simulating the origin of UV-bursts which are thought to originate via high plasma-$\beta$ reconnection in the low solar atmosphere. 
They find that the peak temperature $T_{peak}$ and plasma-$\beta$ are connected (roughly) by a power law of power $-0.3$. 
This means that in high-$\beta$ plasma, magnetic reconnection results in no significant enhancement of the temperature because plasma density is too high, and thus the released energy is distributed over many orders of magnitude more particles than in the solar corona. 
High density has the same dampening effect on outflows from the reconnection region; \cite{tortosa09} found that although outflows did appear in their subsurface reconnection simulations, they were of comparable speed to that of convective flows.

A crucial link between subsurface reconnection and the fast-mode waves necessary to create the I-FIP effect is provided by \cite{kigure10}.
Their 2.5-dimensional magnetohydrodynamical (MHD) simulations  show that component field reconnection between nearly parallel flux tubes can generate Alfv\'en and sound waves in a high plasma--$\beta$ environment.
The Alfv\'en waves originate at the reconnection site and propagate along the reconnected field lines.   
Such waves are mostly incompressible (shear) Alfv\'en waves and they carry approximately 30$\%$ of the total magnetic energy produced by  reconnection in high $\beta$--plasma (compared to $\approx 40 \%$ in low $\beta$--plasma). Sound waves move across the magnetic field and can also carry significant energy.
The conditions simulated by \cite{kigure10} are similar to what is expected with the coalescing flux of a forming light bridge - same polarity flux and high $\beta$--plasma. 

In this work, we hunt for evidence of reconnection below the fractionation layer of the chromosphere in NOAA AR 11504.
We analyse the evolution of the AR using multi-wavelength observations obtained from multiple instruments covering all layers of the solar atmosphere from the photosphere to the corona.
Hinode's EUV Imaging Spectrometer \citep[EIS;][]{culhane07} spectroscopic measurements of the AR's coronal plasma show the presence of I-FIP plasma during episodes of repeated flux emergence principally located above light bridges in both polarities. 
In the context of the ponderomotive force model, I-FIP plasma composition provides indirect evidence of subsurface reconnection \citep{baker19,baker20}.
We infer the large-scale subsurface structure of AR 11504 using the 3D visualization technique of \cite{chintzoglou13} to identify \textit{possible} locations of reconnection among the flux tubes forming the light bridges of the coalescing umbrae.
We then search for other signatures, such as energy release and outflows, of high-$\beta$ reconnection in the solar atmosphere related to the location of the coronal I-FIP plasma. 
 
\begin{figure*}[t!]
\epsscale{1.15}
\plotone{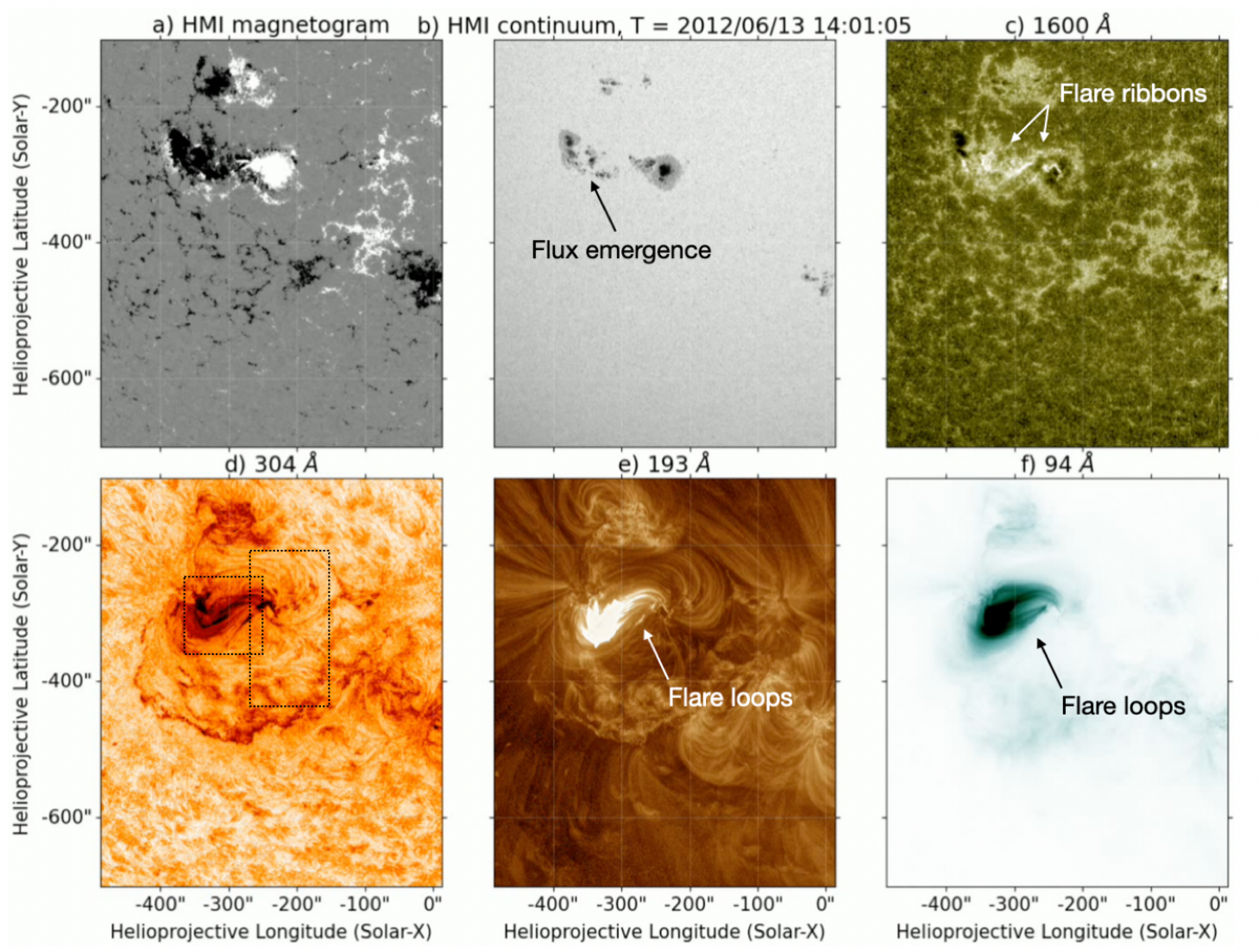}
\caption{SDO/HMI line-of-sight magnetogram (a) and continuum (b), SDO/AIA 1600 $\angstrom$ (c), 304 $\angstrom$ (d), 193 $\angstrom$ (e), and 94 $\angstrom$ (f) at 2012 June 13 14:01 UT.  SDO/AIA 304 $\angstrom$ and 94 $\angstrom$ images are shown in reverse colors.   Magnetogram is saturated at $\pm$200 G.  Dashed boxes in (d) show the approximate EIS FOV on June 13-14 (left) and June 15 (right). 
The animation of the figure contains these six panels covering from 2012/06/13 00:01 UT to 2012/06/15 05:56 UT with a data cadence of 5 min and a total viewing duration of 43 sec.\label{f:sdo}} 
\end{figure*}

\section{AR 11504}\label{s:ar11504}
AR 11504 transited the solar disk from 2012 June 10 to June 20.
Initially, the bipolar AR contained two small sunspots (leading negative/following positive polarities).
However, during the period of June 11-15, significant flux emergence of $\approx 7\times 10^{21}$ Mx occurred in between the two spots in the southeastern part of the AR \citep{james17}.
Distinct fragments of the emerging flux appeared to rotate around the same-polarity (positive) pre-existing sunspot on the western side of the AR \citep[see Table 2 and Figure 8 of][]{james20}. 
The episodic and large scale flux emergence drove the evolution of the AR from a bipolar configuration to a large, highly sheared, complex sunspot group comprising three main sunspots.
Flux emergence ceased on June 15 at about the time that the AR crossed the solar central meridian when its total magnetic flux was $\approx 10^{22}$ Mx \citep[see Figure 2 of][]{james17}, its corrected whole spot area was 1045 millionths of solar hemisphere (MSH; \url{http://fenyi.solarobs.epss.hu}), and its Hale classification was $\beta\gamma\delta$ (\url{http://helio.mssl.ucl.ac.uk/helio-vo/solar_activity/arstats-archive/}).
Figure \ref{f:sdo}(a,b) shows the Solar Dynamics Observatory \citep[SDO;][]{pesnell12} Helioseismic and Magnetic Imager \citep[HMI;][]{scherrer12} photospheric line-of-sight magnetogram and continuum images at 14:01 UT on June 13 during the flux emergence phase.

\begin{figure}[bt!]
\epsscale{1.15}
\plotone{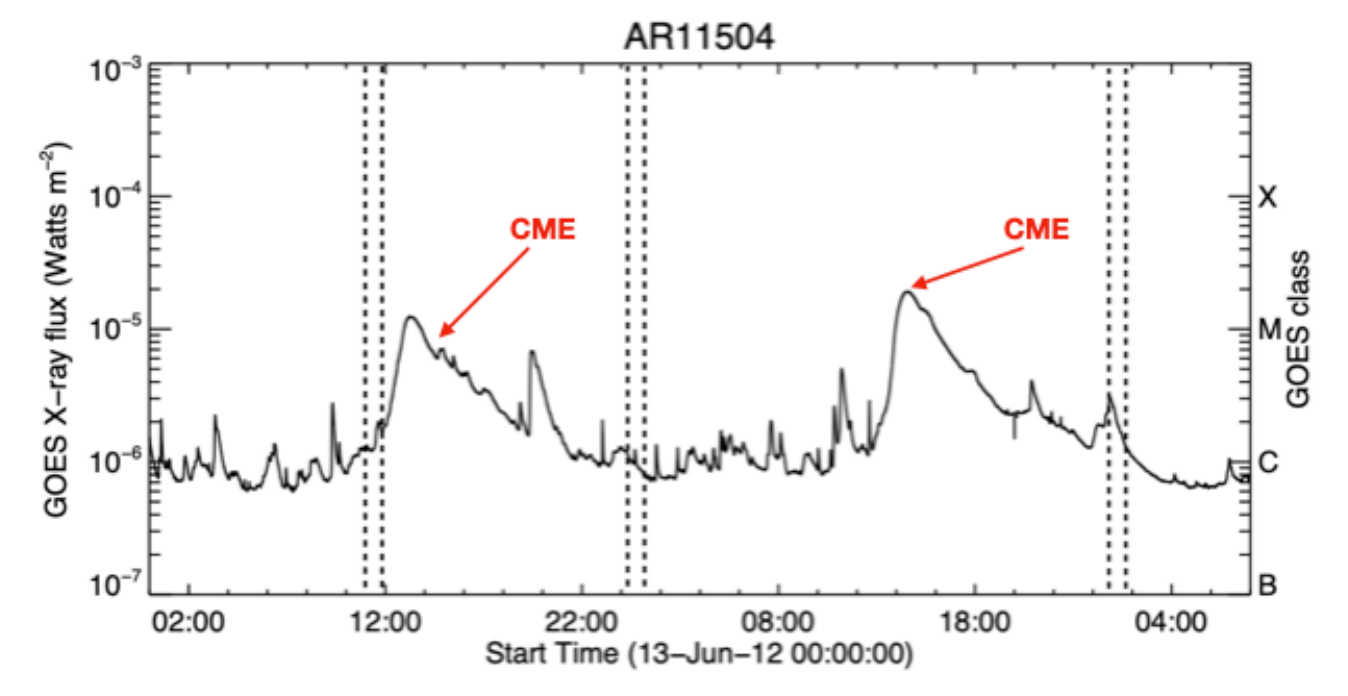}
\caption{GOES X-ray flux overplotted with dashed lines to indicate start/end of Hinode/EIS rasters times (see Figure \ref{f:eis}).  
Arrows indicate approximate times of CMEs \citep{james20}.
\label{f:goes}} 
\end{figure}

The coronal evolution of AR 11504 featured two eruptive events or coronal mass ejections (CMEs) with M-class flares on June 13 and 14. The eruptive event on June 14 produced a gradual solar energetic particle event that was detected by GOES. Both events were preceded by a series of weak, confined homologous flares and their associated ribbons during the formation of the pre-eruptive flux ropes.
These confined flares are interpreted to be signatures of coronal magnetic reconnection associated with the formation of the flux ropes that later erupt as the CMEs   \citep{james17,james20}.
Figure \ref{f:goes} shows the GOES soft X-ray curve covering the period 00:00 UT June 13 to 08:00 UT June 15. 
The approximate timings of the CMEs are indicated by red arrows in the plot.

Figure \ref{f:sdo}(c--f) contains the chromospheric and coronal images of the AR and its surroundings just after the peak of an M1.2 flare and the CME eruption on June 13.
The locations of the flare loops spanning the AR are identified in the coronal/flaring SDO/AIA 193 $\angstrom$ and 94 $\angstrom$ passbands and the ribbons are indicated in the photospheric 1600 $\angstrom$ passband.
The evolution of AR 11504 from 00:00 UT on June 13 to 06:00 UT on June 15 is shown 
in the online animation of Figure \ref{f:sdo}.
Each frame of the movie consists of six panels: (a) SDO/HMI magnetogram, (b) continuum, (c) SDO/AIA 1600 $\angstrom$ (d), reverse-color 304 $\angstrom$, (e) 193 $\angstrom$, (f) and reverse-color 94 $\angstrom$ images with FOV [X, Y] = [500$\arcsec$, 600$\arcsec$].
The data time cadence of the animation is 5 minutes and the viewing duration is 43 seconds.
For a full account of the AR's evolution, its magnetic field configuration, and flux rope formation/eruption, see 
\cite{james17,james18,james20}.

\begin{table}[t!]
	\centering
	\caption{Hinode/EIS study information (upper) and composition diagnostics (lower).  Log T$_{max}$ (K) are from the CHIANTI atomic database \citep{dere97} version 10 \citep{gdz21}.  The FIP values are: Ar = 15.76, Ca = 6.11, Fe = 7.87, S = 10.36 eV.}
	\label{tab_regions}
\begin{tabular}{lcc}
		\hline
		Study ID   & $\#$404 & $\#$420\\
		\hline
		Study name &Atlas$\_$60  & HPW021$\_$\\
        &&VEL$\_$120x512v1\\
	    FOV (arcsec) &120$\times$160 & 120$\times$512\\
        Exposure time (s)& 60& 60\\
        Total raster time (m) & 60& 60\\
        Slit (arcsec) & 2 & 1\\
        Step size (arcsec)& 2 & 2\\
        Raster run time &June 13, 14 & June 15\\  
				\hline
                \hline
        Ion & Wavelength ($\angstrom$)&Log T (K) \\
		\hline
		\ion{Ar}{14} (high FIP)& 194.40  & 6.60\\
		  \ion{Ca}{14} (low FIP)&193.87& 6.65\\
        \ion{Ar}{11} (high FIP)& 188.82& 6.30\\
        \ion{Fe}{14} (low FIP)& 264.79& 6.35\\
        \ion{S}{13} (intermediate FIP)& 256.69& 6.45\\
        \ion{S}{11} (intermediate FIP)& 188.68& 6.35\\        

                \hline
    
			\end{tabular}
	\label{t:EISdetails}
\end{table}

\begin{figure*}[bt!]
\epsscale{1.15}
\plotone{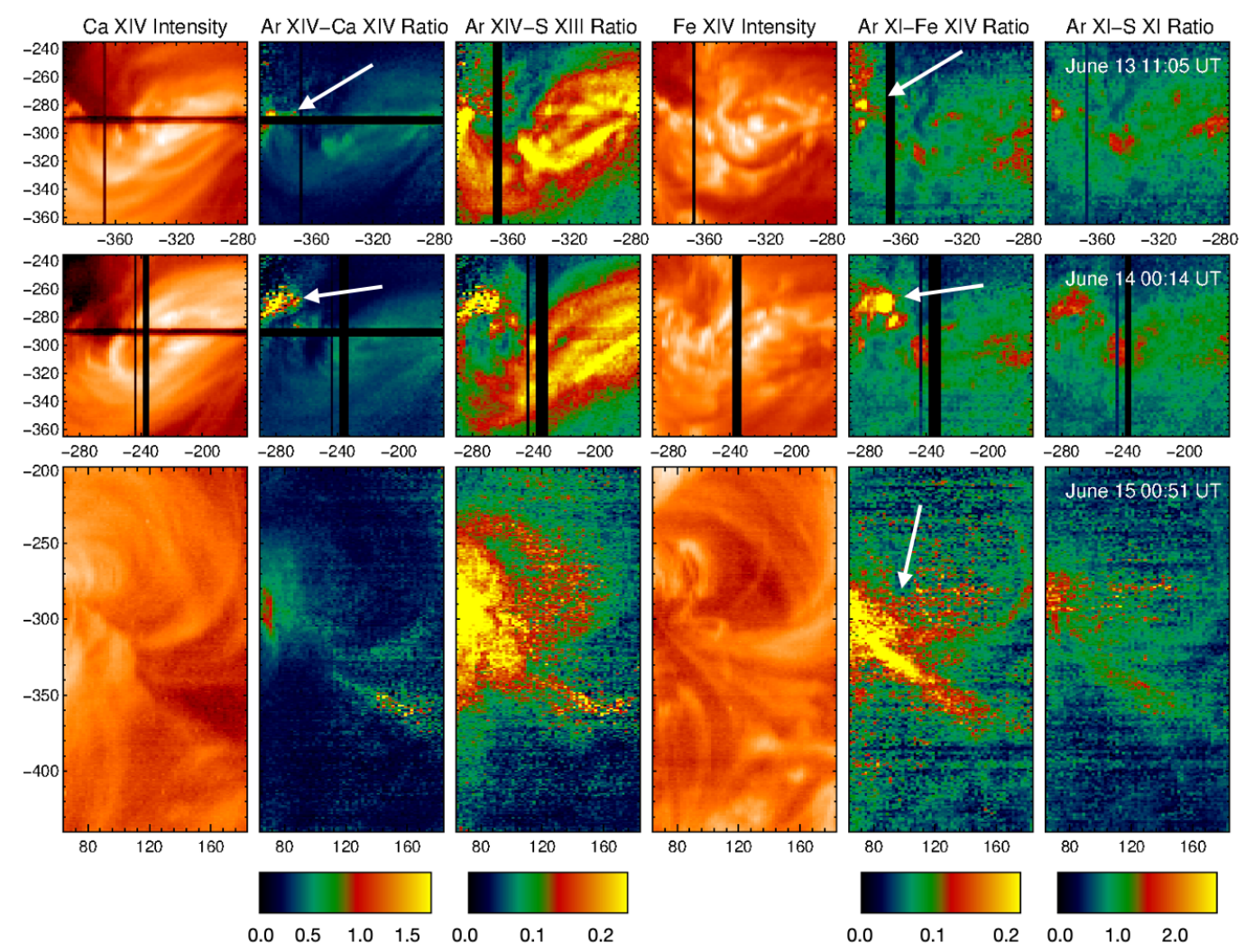}
\caption{Hinode/EIS observations of AR 11504 for 2012 June 13 at 11:05 UT (top), June 14 at 00:14 UT (middle) and June 15 at 00:51 UT (bottom). 
Units are in arcsec. Observations on June 13 and 14 are of the AR's following (negative) polarity and then of the leading (positive) polarity on June 15 (see Figure~\ref{f:locs}). Left to right:   \ion{Ca}{14} 193.87 $\angstrom$ intensity map, \ion{Ar}{14} 194.40  $\angstrom$--\ion{Ca}{14} 193.87  $\angstrom$ and \ion{Ar}{14} 194.40  $\angstrom$--\ion{S}{13} 256.69 $\angstrom$ line intensity ratio maps, \ion{Fe}{14} 264.79 $\angstrom$ intensity map, \ion{Ar}{11} 188.82  $\angstrom$--\ion{Fe}{14} 264.79  $\angstrom$ and \ion{Ar}{11} 188.82  $\angstrom$--\ion{S}{11} 188.68 $\angstrom$ line intensity ratio maps. The color table of the ratios (along the bottom) are saturated at an upper level corresponding to the safe detection of I-FIP, so I-FIP regions are in yellow. Photospheric composition is approximately orange and coronal composition is green/blue, however, these depend on the saturation values used as defined by what is considered to be I-FIP (see end of Section~\ref{s:eis_method} and Appendix).  
(Black stripes are missing data).   
\label{f:eis}} 
\end{figure*}

\section{Hinode/EIS Observations of I-FIP Plasma}\label{s:eis}

\subsection{Method} 
\label{s:eis_method}
Hinode/EIS obtained 26 raster scans of AR 11504 from 2012 June 13--15, three of which contained patches of I-FIP plasma.
Details of the EIS scans where I-FIP plasma was observed are provided in the upper section of Table \ref{t:EISdetails}.
The start/end times of these scans are overplotted on the GOES soft X-ray curve in Figure \ref{f:goes}.
Both EIS studies contain a range of possible emission lines deemed to be appropriate for composition analysis.
The available lines from the full CCD (study name is Atlas$\_$60) were compared to those high and low FIP line pairs with similar emissivity temperature dependencies given in \cite{feldman09}.
The selection criteria for compositional analysis included identifying different diagnostics for probing plasma at multiple temperatures ranging from 0.3 MK to 4 MK and for the line pairs to be well separated in FIP eV. 
Unfortunately, the cooler lines (<1 MK) recommended by \cite{feldman09} were too weak to be used.
There are a number of options for plasma at 1.5--4 MK, including the well-known \ion{Si}{10}--\ion{S}{10} diagnostic for active regions \citep[e.g.][]{brooks11,brooks15}.
However, recent studies have shown that in flaring ARs sulfur often acts like a low-FIP element when used in FIP bias line pairs \citep{doschek16,laming21,to21}.
The FIP of sulfur is 10.36 eV which places it on the cutoff threshold between high- and low-FIP elements.
The dependence of sulfur's behavior on the level of activity in an AR makes it an excellent ion for interpreting plasma composition observations.
Indeed, the behavior of intermediate FIP element of S is proving to be a key composition diagnostic in the solar wind \citep{laming19,parenti21}, with SEPs \citep{brooks21}, and in flares on all scales \citep{doschek17,to21,to24}. 
Its behavior provides insight into the height of the plasma fractionation in the chromosphere. In the upper chromosphere/transition region, S acts like the noble gases with very high FIP and in the low chromosphere, it behaves as a low FIP element.

Because of the observed high-level of activity of AR 11504, other high-FIP elements were considered alongside sulfur.
As was the case in previous studies \citep[e.g.][]{doschek15,doschek16,baker19,baker20,to21,mihailescu23}, high-FIP \ion{Ar}{14} 194.40 $\angstrom$--low-FIP \ion{Ca}{14} 193.87 $\angstrom$ is used to probe plasma of $\approx$ 4 MK.
The behavior and caveats of this composition diagnostic are covered extensively in the literature \citep[e.g.][]{doschek15,doschek16,doschek17,baker19,baker20}.
For $\approx$ 2 MK plasma, high-FIP \ion{Ar}{11} 188.82 $\angstrom$--low-FIP \ion{Fe}{14} 264.79 diagnostic was selected from the recommendations listed in \cite{feldman09}.
The high-FIP and low-FIP elements have similar emissivities within defined temperature domains and they are separated in terms of FIP eV as the noble gas argon has the fourth highest FIP in the periodic table (FIPs: calcium = 6.11 eV, iron = 7.87 eV, and argon = 15.76 eV).
Details of the emission line-pairs used to identify I-FIP plasma in AR 11504 are provided in the lower section of Table \ref{t:EISdetails}.

Level 1 HDF5 data files were downloaded from \url{https://eis.nrl.navy.mil/level1/hdf5/} and calibrated using \cite{warren14}.
Emission lines were fitted using the EIS Python Analysis Code (EISPAC) software \citep{weberg23} and the included fit templates.
Line intensity ratios for \ion{Ca}{14}--\ion{Ar}{14}, \ion{Ar}{11}--\ion{Fe}{14}, \ion{Ar}{14}--\ion{S}{13}, and \ion{Ar}{11}--\ion{S}{11} were calculated for each pixel in the EIS scans.
Figure \ref{f:eis} shows intensity maps of the low-FIP elements and the line intensity ratio maps at 11:05 UT on June 13, 00:14 UT on June 14, and 00:51 UT on June 15.

The line intensity ratio maps used in this investigation may be subject to density and temperature effects.
Ideally, these effects would be removed by using  pixel--to--pixel differential emission measure (DEM) temperature corrections and density measurements to make fully calibrated FIP bias maps \citep[see][]{baker13,brooks15}.
Due to the difficulty in constraining DEM curves at temperatures above those of \ion{Ca}{14} and \ion{Ar}{14} (log T > 6.60 K), fully calibrated FIP bias maps were not a realistic option.
Instead, carefully considered thresholds of what constitutes strong I-FIP plasma for each line pair were determined for the line intensity ratio maps in Figure \ref{f:eis}. 
Each of the maps has been saturated at levels that are considered to be safe estimates of I-FIP plasma taking into account such factors as variations in density and peak temperature in the theoretical ratios for each diagnostic, assumed abundances, cross-detector sensitivity differences, and calibration.
For \ion{Ar}{14}--\ion{Ca}{14} maps strong I-FIP plasma is a ratio {$\geq$ 1.75} and 
for \ion{Ar}{11}--\ion{Fe}{14} maps, it is a ratio {$\geq$0.23}.
A detailed account of what constitutes strong I-FIP plasma for the new diagnostic \ion{Ar}{11}--\ion{Fe}{14} ratio as well as sulfur-based line pairs is provided in the Appendix.

\begin{figure*}[bt!]
\epsscale{1.15}
\plotone{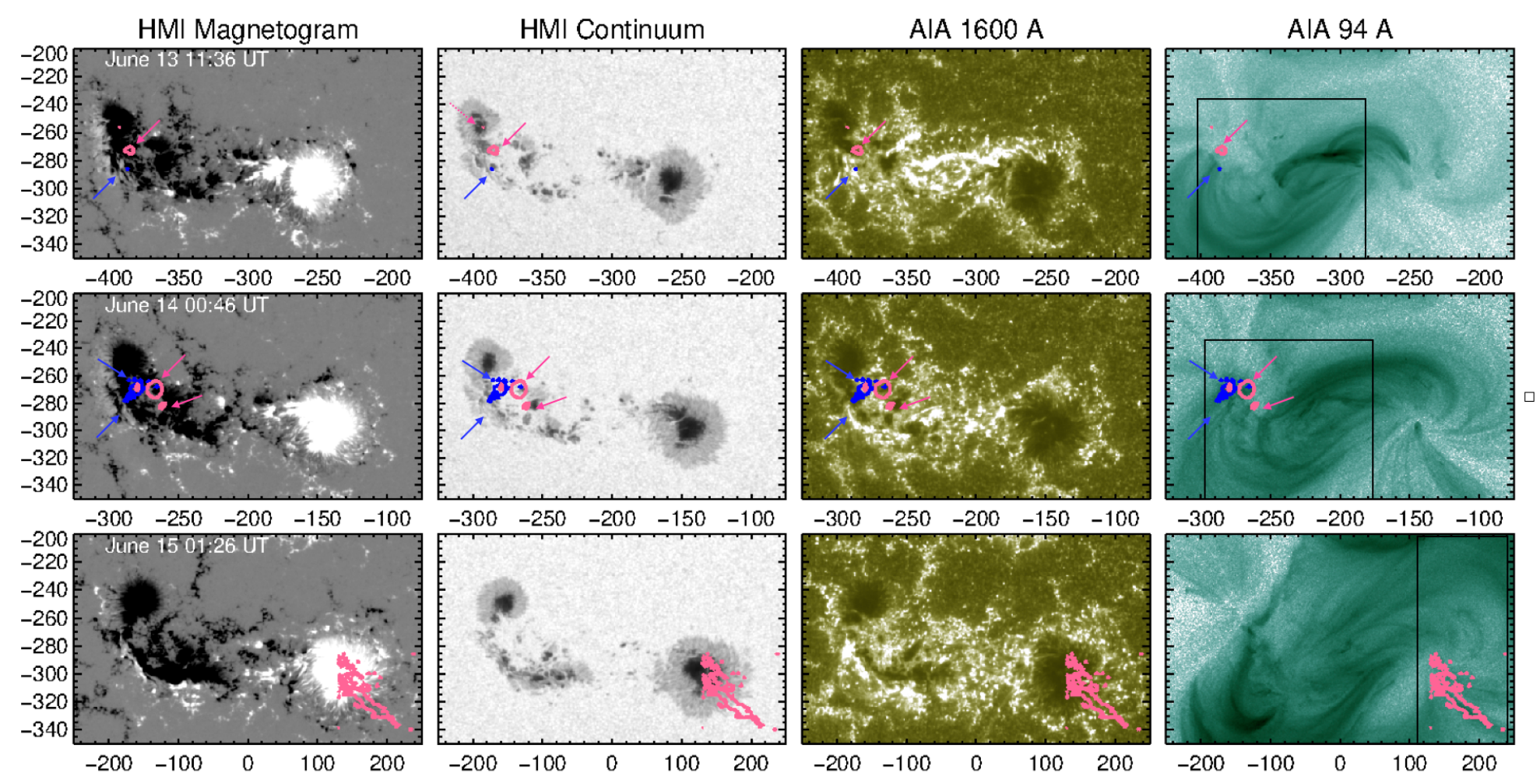}
\caption{SDO/HMI and AIA observations of AR 11504 at 2012 June 13 11:36 UT (top), 2012 June 14 00:46 UT (middle), and 2012 June 15 01:26 UT (bottom). Left to right:   SDO/HMI line-of-sight magnetogram, SDO/HMI continuum, SDO/AIA 1600 $\angstrom$, and SDO/AIA reverse-color 94 $\angstrom$ maps.
Contours (and arrows) are of   I-FIP from the corresponding \ion{Ar}{11} 188.82  $\angstrom$/\ion{Fe}{14} 264.79  $\angstrom$ (pink) and \ion{Ar}{14} 194.40  $\angstrom$/\ion{Ca}{14} 193.87  $\angstrom$ (blue) EIS ratio maps in Figure \ref{f:eis}.  
Dashed pink arrow indicates a few pixels with I-FIP located very close to the light bridge in the northern region of the following sunspot (HMI continuum image on June 13 only).  Black boxes in the right column show the approximate EIS FOV for each raster.
\label{f:locs}} 
\end{figure*}

\subsection{Location of I-FIP Plasma} 

Regions of I-FIP plasma are observed on the eastern side of AR 11504 for June 13 and 14 (two top rows of Figure \ref{f:eis}) in the higher temperature ($\approx$ 4 MK) \ion{Ar}{14}--\ion{Ca}{14} ratio maps (column 2 of Figure \ref{f:eis}). 
The spatial extent of the patch on June 13 is $\approx$ 10$\arcsec$$\times$10$\arcsec$ centered at [X,Y] = [-390$\arcsec$, -290$\arcsec$] and on June 14 is $\approx$ 30$\arcsec$$\times$20$\arcsec$ at [X,Y] = [-275$\arcsec$, -270$\arcsec$].
The lower temperature ($\approx$ 2 MK) \ion{Ar}{11}--\ion{Fe}{14} ratio maps (column 5 of Figure \ref{f:eis}) exhibit a similar behavior though the I-FIP regions are larger compared to the \ion{Ar}{14}--\ion{Ca}{14} ratio maps.

Localized I-FIP plasma is again observed at 2 MK but not at  4 MK at 00:51 UT on June 15, however, the FOV is covering a different region: the western side of AR 11504.
The plasma has evolved to photospheric composition in the \ion{Ar}{14}--\ion{Ca}{14} ratio map. 
Interestingly, in the \ion{Ar}{11}--\ion{Fe}{14} ratio map, I-FIP plasma appears to extend to the south/southwest along loops rooted in the light bridge containing the I-FIP patch centered at [X,Y] = [85$\arcsec$, -295$\arcsec$].

All regions of I-FIP plasma are surrounded by a ring of weak I-FIP/photospheric plasma embedded in the dominant coronal composition plasma.
The precise extent of the I-FIP plasma and the boundaries where the plasma transitions to photospheric and then to coronal composition  depend on the definition of I-FIP used. 
It is important to note that I-FIP plasma may have occurred at locations and times other than in the three scans presented here, however, due to a number of factors, EIS did not detect it.
Some of these factors include: the limited FOV of EIS, telemetry constraints, pointing changes, and the lack of composition diagnostics.

Columns 3 and 6 of Figure \ref{f:eis} show line intensity ratio maps where intermediate-FIP sulfur ions are used instead of the low-FIP calcium and iron ions.
In the case of the \ion{Ar}{14}--\ion{S}{13} diagnostic, sulfur behaves like a low-FIP element on June 13 and 14.
This is the case where I-FIP plasma is detected in the \ion{Ar}{14}--\ion{Ca}{14} maps and  throughout the core of the AR (cf. columns 2 and 3 of Figure \ref{f:eis}).
Sulfur is less fractionated in terms of both degree and spatial extent when comparing the lower temperature \ion{Ar}{11}--\ion{S}{11} and the higher temperature \ion{Ar}{14}--\ion{S}{13} ratio maps (columns 3 and 6).

The specific locations of the I-FIP plasma within AR 11504 are shown in Figure \ref{f:locs} which displays SDO/HMI magnetic field, HMI continuum, SDO/AIA 1600 $\angstrom$, and reverse-color AIA 94 $\angstrom$ maps timed at the mid-points of the EIS scans in Figure \ref{f:eis}.
The contours of I-FIP plasma from the EIS scans are overplotted on the SDO/HMI and AIA maps. 
Pink/blue contours and arrows correspond to I-FIP patches in the \ion{Ar}{11}--\ion{Fe}{14} / \ion{Ar}{14}--\ion{Ca}{14} line ratio maps.

On the eastern side of AR 11504, patches of I-FIP plasma are observed in the negative  polarity fragments forming the southern of the two following sunspots.
This is the location of the large scale flux emergence that occurred from June 11--15.
A tiny patch is also found at the western edge of the major light bridge of the northern sunspot (indicated by the dashed pink arrow in the top row of Figure \ref{f:locs}).
In the leading (positive) sunspot, the I-FIP plasma lies above the strong light bridge separating the large pre-existing spot and a smaller spot that results from new flux emergence. 
The above analysis of the photospheric evolution suggests that the spots may represents large flux tubes that are actually interacting below the visible surface.

Both sites are associated with the footpoints of the flaring loops spanning the AR and repeated flare ribbon crossings of the coalescing leading and following sunspots. 
An example associated with the CME on June 13 is shown in Figure \ref{f:sdo} where the flare ribbons are cospatial with the observed I-FIP plasma in the top panels of Figure \ref{f:eis}. Although the EIS scan was taken mostly prior to the CME, the 60-min scan time and the W--E scanning direction mean that the I-FIP patch on the eastern edge of the raster was taken at about 12 UT on the 13th, at the start of the flare/CME event. 
Repeated flaring on all scales means that there are frequent occurrences of flare ribbons located above the forming sunspots on either side of the AR
(See the SDO/AIA 1600 $\angstrom$ panel in the animation of Figure \ref{f:sdo}).

\subsection{Relating Hinode/EIS Observations to the Ponderomotive Force Fractionation Mechanism} 
\cite{laming15} proposed the ponderomotive force associated with Alfv\'enic waves as the  fractionation mechanism that separates ions from neutrals in the chromosphere leading to the well-known FIP effect \citep[see also ][]{laming04,laming09,laming12}.
The ponderomotive acceleration is proportional to the gradient of  $\delta$E$^{2}$/B$^{2}$, where $\delta$E is the wave electric field and $B$ is the ambient magnetic field.
Since the magnetic field strength usually decreases with height, the upward gradient of $\delta$E$^{2}$/B$^{2}$
is generally positive which produces  the FIP effect; a negative gradient results in I-FIP plasma.

In the case of the I-FIP effect, fast mode waves originating at/below the photosphere undergo total internal reflection and refract back downwards in the lower chromosphere where the Alfv\'en speed is increasing with height \citep[i.e. the wave amplitude increases with decreasing density;][]{laming21}.
It is this total internal reflection of waves that provides the downward pointing ponderomotive force in the model.
The force acts only on ions, thereby depleting the plasma of the easy to ionize low FIP elements and enhancing the \textit{relative} abundance of high FIP elements to create the I-FIP effect.
The I-FIP effect plasma is then transported to the corona by chromospheric evaporation during flaring.

Whether the waves are generated below the chromosphere, at the photosphere, or subsurface (meaning subphotospheric) is not critical to the ponderomotive force model.
The most important factor is that once the waves propagate above the plasma $\beta$ = 1 layer, they reflect/refract quickly back downwards and generate I-FIP fractionation.
In the highly complex chromosphere, it is likely that different geometries would produce I-FIP effect plasma \cite[e.g.][]{martinez23}.

Assuming a sufficiently strong source of fast mode waves from below where high-FIP and low-FIP ions are fractionated in the chromosphere, the simulations of \cite{laming21} show that the degree of I-FIP fractionation primarily depends on: 1. how much the magnetic field expands from the photosphere to the corona (expressed as $B$$_{cor}$/$B$$_{photo}$), and 2.  the ionization balance of key elements in the lower chromosphere.
For the case of minimal expansion in the magnetic field ($B$$_{cor}$/$B$$_{photo}$ = 0.7),  low FIP elements, including Fe and Ca, and intermediate FIP S were depleted low in the chromosphere just above the plasma $\beta$ = 1 layer set at 350 km \citep[see Figure 3 (e,f) of ][]{laming21}.
This is the height at which the perpendicularly propagating fast mode waves were undergoing total internal reflection and creating the negative ponderomotive acceleration.
High FIP Ar was  unaffected by the downward directed ponderomotive force as it is essentially neutral in the model chromosphere.
The Hinode/EIS \ion{Ar}{14}--\ion{Ca}{14} and \ion{Ar}{11}--\ion{Fe}{14} line intensity ratio maps are in full agreement with the simulations.
The low FIP \ion{Ca}{14} and \ion{Fe}{14} are depleted relative to their high FIP Ar ion counterparts as is evident in Figure \ref{f:eis}.
Like Ca and Fe, S is also largely depleted at the higher temperature of $\approx$ 4 MK as shown in the \ion{Ar}{14} -- \ion{S}{13} line ratio maps.
However, S appears to be only partially depleted relative to the lower temperature of $\approx$ 2 MK.
The fractionation of S depends on collisional interactions with the background gas of neutral H, as opposed to ionized H higher up in the chromosphere.

\begin{figure*}[bt!]
\epsscale{1.15}
\plotone{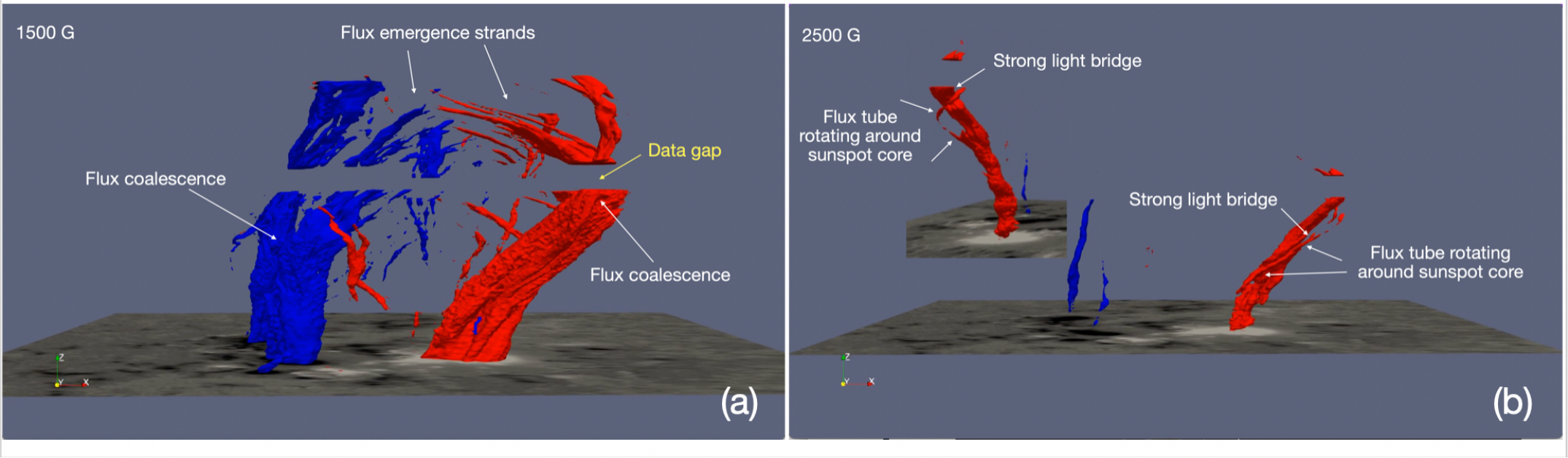}
\caption{3D visualization of the subsurface magnetic structure of AR 11504 constructed from stacking contours at  1500 G (a) and 2500 G (b) from the B$_{r}$ component of SHARP vector magnetograms using the method of \cite{chintzoglou13}.  
Time runs from 17:00 UT June 9 (top) to 13:00 UT June 20 (bottom) of each image.  
Inset in (b) shows the rotating flux tube around the sunspot core of the positive polarity from a different viewing angle.
Key locations for subsurface reconnection to take place are: 1.  flux coalescence in the following (blue) polarity and 2.  at the strong light bridge in the leading polarity (positive, see Section~\ref{s:search}).
The animation of the figure shows the orbital view of the 3D structure first at 1500 G then at 2500 G.  The total viewing time is 48 sec.
\label{f:3d}} 
\end{figure*}

\begin{figure*}[bt!]
\epsscale{1.15}
\plotone{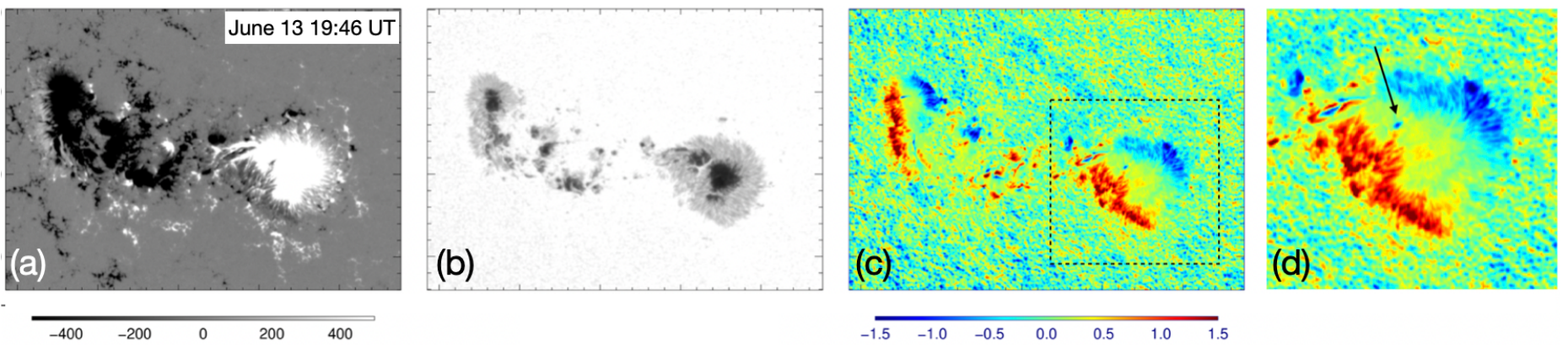}
\plotone{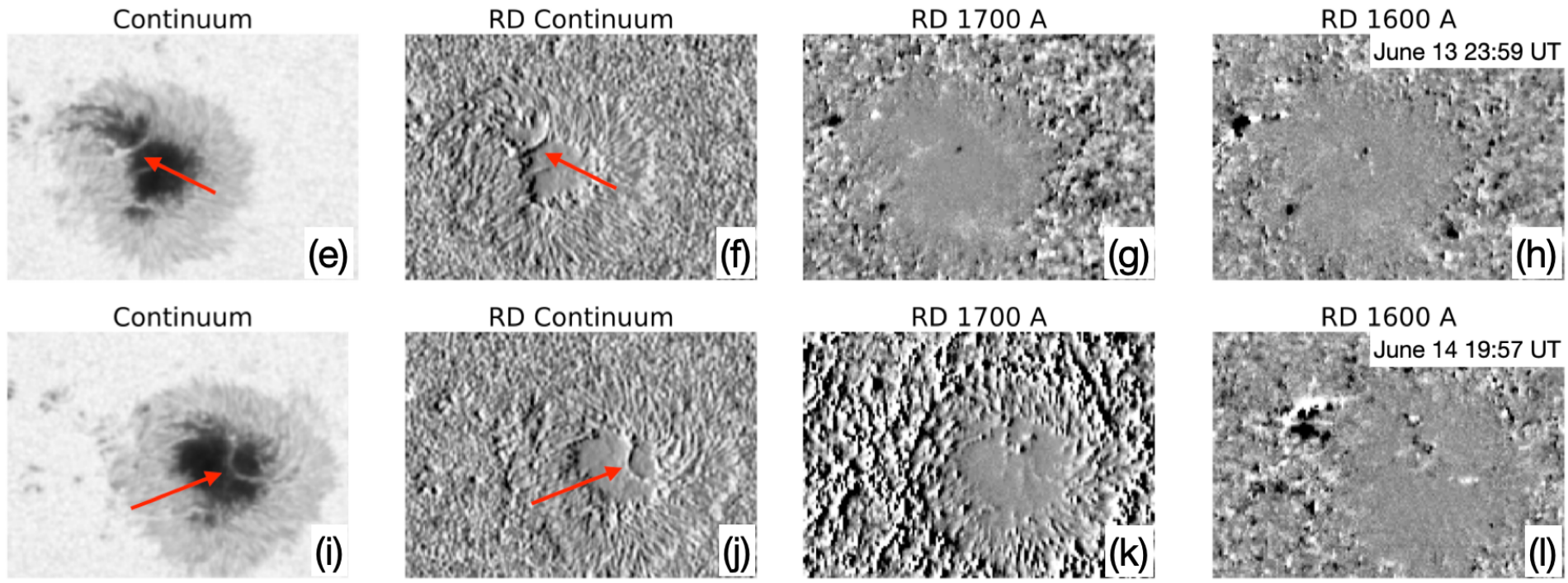}
\caption{Photospheric observations.  
HMI LOS magnetogram saturated at $\pm$500 G, (a), continuum intensity (b), Dopplergram (c) and zoomed Dopplergram of boxed region (d).The color scales are in Gauss for the magnetic field and in km s$^{-1}$ for the velocity. 
In panel (d), the black arrow indicates strong upflows at the light bridge. 
The two lower rows show SDO/HMI continuum, running difference (RD) continuum, RD AIA 1700 $\angstrom$ and RD 1600 $\angstrom$ images at 23:59 UT on June 13 (e-h) and 19:57 UT on June 14 (i-l).  
Red arrows point out the ridges of brightenings along the light bridge formed by the coalescence of the leading polarity sunspot.  
These ridges are not evident in the cospatial/cotemporal  SDO/AIA images of the upper photosphere and transition region (panels g,h,k,l).  
The cadence of the running difference movie (not shown), from which these images were taken, is 45 sec.
The animation of the figure contains panels (a,b,c) covering from
2012/06/12 23:58 UT to 2012/06/15 11:46 UT with a data cadence of 12 min and a total viewing duration of 18 sec.
\label{f:photos}} 
\end{figure*}

The I-FIP effect plasma observed by EIS in AR 11504 is consistent with the predictions of the ponderomotive force fractionation model, suggesting that there is a source of fast-mode waves originating from below the fractionation region in the chromosphere.
In fact, with its magnetic field strength measured as  $>$2700 G \citep{pal18}, the source of the fast-mode waves is plausibly at/below the photosphere of AR 11504 since the plasma $\beta$ = 1 layer is located lower in regions of strong magnetic field \citep{gary01,avrett15,bourdin17}. 
A candidate for the origin of the required fast mode waves is subchromospheric reconnection.
Such waves are spatially localized in magnetic flux tubes and are more likely to have sufficient wave amplitude to cause the fractionation, in contrast to much lower amplitude acoustic waves  excited by convective motions that are present everywhere in the chromosphere \citep{laming21}. 

Other models/simulations also invoke the ponderomotive force as the main mechanism for fractionation in the chromosphere/transition region  \citep[e.g.][]{reville21,martinez23}.
They focus mainly on the FIP effect rather than the I-FIP effect, however, and on the whole the EIS observations of AR 11504 do not appear to contradict their results.

\section{Search for conditions and signatures of subchromospheric reconnection} \label{s:search}
As described in Section \ref{s:intro}, magnetic reconnection is expected to take place low-down in the solar atmosphere and even below the photosphere in high-$\beta$ plasma, where non-parallel magnetic fields have to be brought together by plasma motions. 
Simulations show that current sheets can form and reconnection can take place under such conditions, leading to energy release and outflows (see Section \ref{s:intro}). 
However, due to the high densities, the resulting increase in temperature is weak, and the outflow speed is of the order of the typical convective speed, so the reconnection consequences may be at the limit of what can be detected. Nevertheless, in the following we make an attempt to search for signatures of heating and photospheric upflows at the places where flux emergence is bringing magnetic flux tubes together, e.g. in merging umbrae, and in particular when new flux is approaching and wrapping around old flux, forming a prominent light bridge, as we observe in the big leading spot of this AR. 
Light bridges are interfaces  between magnetic flux tubes, and in case there is a non-zero angle between the same-polarity magnetic fields being pushed together by ongoing flux emergence, they are likely locations for reconnection below the fractionation region. 
There are light bridges in both the leading and following regions of this AR, and the observed I-FIP plasma patches are indeed located in their vicinity. 
However, the following polarity light bridges in coalescing umbrae are smaller and less long-lived than in the big leading spot, so we mainly concentrate our analysis on the latter, where we can clearly follow the formation and evolution of a long light bridge with patches of I-FIP plasma above it (due to EIS pointing choices on June 15).
In the following, we look for conditions and consequences of reconnection there.

\subsection{Condition: Non-parallel Flux Tubes Pushed Together}
\label{s:3d}

The subsurface structure of AR 11504 can be inferred from observations using the image-stacking technique of \cite{chintzoglou13}.
An abridged description of the image-stacking method is as follows.
A 3D data cube is constructed from SDO/HMI Space-weather HMI Active Region Patches \citep[SHARPs;][]{bobra14} vector magnetograms. 
The starting time, t$_{0}$, is at the top of the data cube and the 2D magnetograms forming the X--Y plane are added at progressively lower `heights' in the Z-direction.
Contours of radial magnetic field, B$_{r}$, from each magnetogram are then stacked along the time dimension.
The overall picture from this analysis is an approximation of the sub-photospheric 3D magnetic flux structure at time t$\_$0 as it would result from a purely solid-body (kinematic) emergence. 
The solid-body assumption means that the effects of the turbulent convection zone and emergence through the photosphere are not captured.
Caution is therefore required when interpreting the 3D structure, especially in the case of weak field.
Nevertheless, this technique can help to visualize the main characteristics of the evolution of an active region 
For a full account of the method, see \cite{chintzoglou13} and references therein.

The visualization of the 3D structure of AR 11504 from 2012 June 9--20 is displayed in Figure \ref{f:3d}(a,b) and the corresponding online animation.
The movie contains an orbital view of the volume containing the respective stacks of B$_{r}$ magnetic contours overplotted on HMI vector magnetograms on June 20. 
Red/blue represent positive/negative magnetic field.
It lasts for 48 sec.

Contour levels of 1500 G and 2500 G (Figure \ref{f:3d}(a,b)) were selected to highlight different features of the emerging structure.
The B$_{r}$ stacking of 2500 G contours emphasizes the larger scale magnetic structure, the so-called `trunk', whereas, the stacking of 1500 G contours provides a sense of the smaller `branches' of the structure \citep{chintzoglou13}.

Light bridges formed as individual flux tubes rotated around the core of the leading sunspot.
One of the main light bridges is indicated in the red B$_{r}$ stack in the upper right panel of Figure \ref{f:3d}(b)  and corresponds to the light bridge of the positive polarity evident in the continuum image (Figure \ref{f:photos}(b)).
I-FIP is observed in the EIS line ratio maps at 00:51 UT on June 15 precisely above this light bridge separating the rotating flux tubes (Figure \ref{f:locs}, bottom row).
Furthermore, the loops containing I-FIP plasma appear to be rooted very close to the light bridge (Figure \ref{f:eis}, bottom row).

A more detailed view of the topology is provided by the stack of 1500 G contours in the 3D visualization.
The region of large scale flux emergence mentioned in Section \ref{s:ar11504} is highlighted in the negative/following polarity (blue) in Figure \ref{f:3d}(a). 
Before the data gap on June 13, numerous individual strands of emerging flux are clearly identifiable in the following sunspot(s) unlike in the more coherent leading one.
After the data gap, the flux emergence has stopped in the south/southeastern  section of the AR and the flux has coalesced to form two coherent (blue, negative polarity) sunspots.
This is the location where I-FIP was observed on June 13 and 14 while flux emergence/coalescence was taking place (Figure \ref{f:locs}, top and middle rows). 

The 3D structure in Figure \ref{f:3d}(a,b) can be interpreted as an approximation of the subphotospheric structure that, by emerging rigidly through the photosphere, would produce the observed photospheric evolution. 
Within the limits of such a visualization, one can still infer the complex dynamical evolution of the coalescing magnetic flux strands that finally result in the observed AR's $\beta\gamma\delta$ Hale classification. 
For instance, the red magnetic flux tubes that are pushed together to form the leading polarity have the same, dominant radial component, but component-reconnection can still occur between the horizontal component in the volume between them.

\begin{figure*}[bt!]
\epsscale{1.15}
\plotone{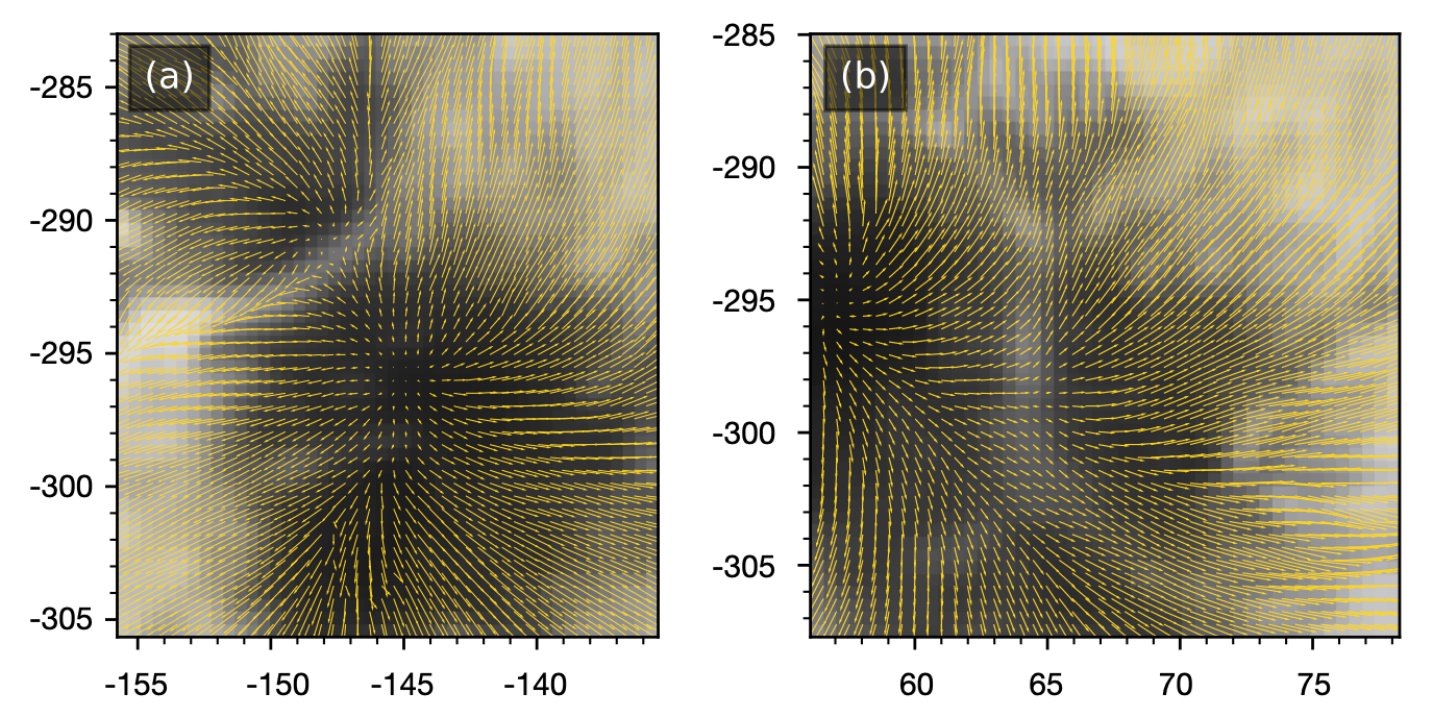}
\caption{White light continuum images of the leading positive sunspot on (a) 2012 June 14 at 00:58 UT and (b) 2012 June 15 at 00:58 UT with arrows that represent the strength and direction of the horizontal magnetic field. In panel (a), the horizontal field on either side of the light bridge (approx $x=-147''$, $y=-292''$) has small antiparallel components with very weak horizontal field in the center of the light bridge. In panel (b), the horizontal field is parallel on either side of the light bridge, but the horizontal field is weaker in the middle of the light bridge (approx $x=64''$, $y=-299''$).
\label{f:horizontal}} 
\end{figure*}

\cite{baker19,baker20} posited that strands of sheared field coalescing during the formation of sunspot umbrae suggests the possibility of subsurface component reconnection.
Component reconnection requires a nonzero component of anti-parallel field.
Such a magnetic field configuration appears in the subsurface structure of leading and following sunspots.
Two key areas are (i)  the region of large scale flux emergence in the following polarity, and (ii)  the flux tube rotating around the core of the leading spot that forms a strong light bridge in between them.
I-FIP plasma is observed in the low corona that is plausibly connected to subsurface magnetic field in these locations. 
These light bridges are locations where the conditions are possible for subphotospheric reconnection to take place.
However, it is important to note that the 3D visualization of the subsurface structure provides a road map of potential locations but no information as to whether reconnection has taken place below the photosphere.

Examining the horizontal magnetic field in the leading sunspot, we observe small antiparallel field components on either side of the light bridge on June 14 
(see Figure \ref{f:horizontal}(a)). 
On June 15, closer in time to the IFIP observation in the leading spot, the horizontal field direction is parallel on either side of the light bridge (see Figure \ref{f:horizontal}(b)). 
However, the horizontal field strength is significantly weaker in the center of the light bridge than it is on either side of the structure which is typical in light bridges (see references in Section \ref{s:intro}).
In AR 11504, the horizontal field, B$_{tr}$ is $\sim$200 G in the light bridge versus $\sim$1500 G in the surrounding regions.

It is interesting to note that if the magnetic field we observe in Figure \ref{f:3d}(a,b) is comparable to the field destroyed by reconnection, we can estimate whether there is enough energy for I-FIP fractionation to result from reconnection.
For example, when the magnetic energy at 1500 G is converted into waves at a plasma density of 10$^{17}$ cm$^{-3}$, then the wave amplitude is approximately 10 km s$^{-1}$.
This level of wave amplitude is of the right order to produce the I-FIP effect according to the ponderomotive force fractionation model.
Of course, Figure \ref{f:3d}(a,b) shows the magnetic field that survives rather than what is destroyed by reconnection.
However, the numbers, such as we can constrain them, all support the basic idea of reconnection generating I-FIP fractionation in AR 11504.

\subsection{Consequences: Reconnection Outflow and Heating} \label{s:v_field}
We search for signatures of subchromospheric reconnection outflows in SDO/HMI (photospheric) Dopplergrams. The online animation of Figure \ref{f:photos}(a--c) shows the formation of the light bridge as the pre-existing positive-polarity leading spot is approached by newly emerging positive-polarity flux concentrations.  
Each frame contains SDO/HMI LOS magnetograms saturated at $\pm$500 G, continuum images and Dopplergrams showing photospheric upflows/downflows in blue/red.
The FOV [X, Y] = [250$\arcsec$$\times$170$\arcsec$].
The time period covered is from 2012 June 12 23:59 UT to June 15 05:47 UT at a time cadence of 12 minutes. 

Of particular importance for locating reconnection outflows (appearing as upflows), in the right panel of the movie and Figure \ref{f:photos}(c), SDO/HMI hmi.V$\_$720s Dopplergrams are shown. 
They were downloaded from the the Joint Science Operations Center (JSOC) website (\url{http://jsoc.stanford.edu}).
Dopplergram velocities were  calibrated taking into account the orbital motion of SDO/HMI and the radial velocity of the Sun using the HMI ring diagrams available from JSOC. 
The mean Doppler velocity of each map is subtracted from all pixels within the map \citep{couvidat16,murabito21}.

The Doppler maps show effects of the Evershed flow in the penumbra of the main sunspots. We also see downflows over the leading spot's umbra, with a faint trace of the light bridge, where the velocities are mainly around zero. 
In these Dopplergrams, we are looking for blue-shifted episodic upflow events along the light bridge as potential tracers of subchromospheric reconnection outflows. 
Previous studies \cite[e.g.][]{leka97,toriumi15a} reported upflow velocities of a few $0.1$ to 1.5 km s$^{-1}$ along light bridges, which have a weak, more horizontal field than the neighboring umbrae. 
Furthermore, convective intrusion driving magnetic reconnection at lower altitudes is also a possibility \citep{toriumi15b}.
There is only one clear upflow event seen in the movie along the light bridge in AR 11504.
This upflow is shown in Figure \ref{f:photos}(c,d). 
It is seen for about 2 hours (on June 13 between 19:10--21:22 UT), around the time of the formation of the light bridge, when EIS was observing the following polarity region of AR 11504. 
Unfortunately, there is no co-temporal corroborating I-FIP observation of this potential reconnection outflow event. 

Next, we search for signatures of heating potentially resulting from subchromospheric reconnection in running-difference (RD) movies (not shown) of co-aligned SDO/HMI continuum, SDO/AIA 1700 and 1600 \AA~ images, which represent increasing height from the photosphere. 
In Figure \ref{f:photos}(e--l), we show snapshots of SDO/HMI continuum and AIA 1700 and 1600 \AA~ running-difference images. 
(See Sunpy documentation for making RD images at \url{https://docs.sunpy.org/en/stable/generated/gallery/index.html}).
We find that in RD continuum the light bridge is clearly seen as a black-and-white patterned sharp feature, in RD at 1700 \AA~ it is still discernible, but in RD 1600 \AA~ it is not evident at all. 
As the atmospheric layers these images sample are increasing in height from continuum to 1600 \AA, it gives a hint of low-altitude weak energy release being present along the light bridge in the photosphere, which then wanes in the lower atmosphere. 
The light bridge is particularly prominent at its formation on June 13 and increased brightness seems to be present in the vicinity of the upflow event observed in the Dopplergrams in Figure \ref{f:photos}(d), however, the heating at the formation height of 1600 $\angstrom$ does not appear to be as concentrated as the blue-shifted upflows there.

\section{Discussion}\label{s:disc}

During a period of large-scale flux emergence in the highly complex and active AR 11504, the I-FIP effect was observed in plasma at $\approx$ 2 MK and $\approx$ 4 MK along light bridges in the coalescing umbrae of both AR polarities.
Consistent with previous studies \citep{baker19,baker20}, the regions of I-FIP plasma were localized at the footpoints of flare loops in so-called monster ARs of high complexity (Hale classification of $\beta\gamma\delta$), strong total magnetic flux (>10$^{22}$ Mx), and large sunspot area (>1000 MSH).

Interestingly, on June 15, the I-FIP plasma area also extends at least 50$\arcsec$ along loops rooted in the vicinity of the strong light bridge of the leading spot (Figure \ref{f:eis}, bottom row).  
These are the first known observations showing I-FIP plasma at a distance away from flare loop footpoints. 
The timing of the Hinode/EIS observation coincided with the peak of a C 3.0 class flare that occurred during the extended decay phase of an M1.9 class flare  (see Figure \ref{f:goes}), making it possible to observe the I-FIP plasma along the loops.
As noted in \cite{baker19,baker20}, I-FIP plasma is observed when it is evaporated into flare loops that cross sunspot umbrae, however, this can not be directly demonstrated on June 15 as EIS FOV does not fully cover the footpoints of the loops containing the I-FIP plasma.
Nevertheless, on June 14 at 00:14 UT, the circumstances are similar to those on June 15.
The EIS raster scan covered the negative polarity which hosted multiple small flaring  events during the extended decay phase of an M1.2 class flare.
The I-FIP region (see Figure \ref{f:eis}, middle row) is dominated by upflows characteristic of chromospheric evaporation.
Doppler velocities are in the range = [-100, -20] km s$^{-1}$ with a mean value of -47 km s$^{-1}$, much greater than upflow velocities found in quiescent active regions \citep{tian21}.
It is therefore plausible that the pristine I-FIP plasma was evaporated into the loops during the C3.0 class flare on June 15 and was observed before fully mixing with the low-FIP enhanced (i.e., high FIP bias) plasma already contained in the loops.
The orange ring of weak I-FIP/photospheric plasma surrounding the elongated region of I-FIP is an indication of plasma mixing in the loops (see the \ion{Ar}{11}--\ion{Fe}{14} line ratio map in the  bottom panel of Figure \ref{f:eis}).  
The same feature is notable in all of the line ratio maps shown here as well as in those of other studies of spatially resolved plasma composition during flares \citep[e.g.][]{baker19,baker20} where the orange ring encircles I-FIP patches at loop footpoints (rather than along the loop).
In AR 11504, the I-FIP effect is present in an extended region away from loop footpoint, suggesting that I-FIP plasma is not restricted to small regions.
Though on a much smaller scale, this is consistent with the findings of \cite{katsuda20} where the I-FIP effect made up the bulk of the plasma in four giant X-class flares. 

Though the presence of I-FIP plasma in the corona is not direct evidence of subchromospheric reconnection, it nonetheless provides solid clues about the likely conditions low down in the solar atmosphere/photosphere.
I-FIP effect results when there are sufficiently large fast-mode wave trains generated below the equipartition layer in order to produce the required downward ponderomotive force \citep{laming21}.
This does not appear to occur under all flaring conditions in ARs, otherwise, I-FIP plasma would presumably be observed more frequently over a variety of regions.
So the presence of I-FIP plasma tells us that there are more fast-mode waves generated locally rather than globally beneath the surface.
Subchromospheric reconnection is 
likely to be the source of the required wave trains especially in strong sunspots. 
The fact that sulfur acts like a low-FIP element in AR 11504 further supports the argument that the waves are created below the fractionation region in the chromosphere, and that the fractionation occurs low in the chromosphere.
The few active regions where I-FIP effect plasma has been detected are highly complex with unusually strong coalescing sunspots.
In such strong sunspots, the plasma $\beta$ = 1 layer is likely to be low down in the photosphere or even subphotosphere \citep{carlsson19}, leaving open the possibility of reconnection happening below the solar surface.
However, reconnection within light bridges is likely to be in/just above the photosphere.
Therefore, highly localized regions of I-FIP plasma in the corona indicate possible locations of such low-down reconnection and its consequences in the corona.

In this paper, we were searching for observational evidence --other than I-FIP plasma observed in the corona, but in tandem with it-- of subchromospheric/high-$\beta$ plasma magnetic reconnection. 
We found that in AR 11504 the \textit{conditions} for subchromospheric reconnection have been met during the 2nd major episode of flux emergence, between June 13--15, when new flux approached and merged with pre-existing sunspots and smaller magnetic concentrations were also merging. 
In particular, in the leading-polarity spot, the new flux wrapped around the pre-existing flux (see Figure \ref{f:3d}), implying the possibility of significant non-parallel orientation between the two large positive-polarity concentrations, likely enabling magnetic reconnection along their interface. 
\cite{james20} found that the rotation rate of the new flux around the core of the sunspot  ($\approx$ 2.6$^{\circ}$ hour$^{-1}$) and the total orbit (186$^{\circ}$) peaked during the period when I-FIP plasma was observed in AR 11504.
A prominent light bridge signified the interface between the two magnetic flux tubes/sunspot umbrae, so this was the main target for our search of signatures of episodic heating and reconnection outflows (potentially observed as upflows). 

Light bridges usually have heating and upflow events \citep{toriumi15b}, but these can be simply attributable to effects of magnetic reconnection driven by magnetoconvection \citep{toriumi15a}. 
As \cite{tortosa09} has shown, the reconnection outflows are likely to be comparable in speed with that of the convective motions. 
We did find an upflow event lasting for about two hours along the bright newly formed light bridge, but we did not have co-temporal plasma composition information to corroborate that it signified subchromospheric/high-$\beta$ plasma reconnection.
It is likely that such subtle effects, upflow and heating evidence, are at the limit of detectability in HMI observations. 
From the simulation results of \citet{toriumi15a,toriumi15b}, it is not surprising that we found only weak evidence of enduring photospheric heating along the light bridge in running-difference continuum images  whose effects diminish with increasing height, and become unobservable in AIA 1700 and 1600 \AA~ images. 
Though the observable signatures of reconnection related to light bridges might be difficult to detect, it is clear that light bridges are natural locations for reconnection to occur at/just above the photosphere and thus a potential source of fast mode waves that can generate I-FIP plasma. 

\section{Conclusions}\label{s:concl}
Direct observational evidence of subchromospheric reconnection remains elusive, however, a preponderance of indirect evidence has contributed to our understanding of the conditions required for it to take place and consequences of the processes associated with it.
In AR 11504, there were signs of subtle heating and weak short-lived reconnection outflows observed in the photosphere as upflows. 
We would expect these pieces of evidence of reconnection to be linked to the presence of the I-FIP effect, although, the observations were not cotemporal.
However, they were all localized in, or at least nearby, light bridges / in between magnetic flux tubes at the photospheric level.

Among the new directions to pursue is the analysis of changes in sunspot motions and morphology due to subchromospheric reconnection, employing techniques such as the 3D visualization of AR subsurface topology used here.
We anticipate that other indications of reconnection in high-$\beta$ plasma will become more evident with the high resolution observations provided by DKIST, Solar Orbiter, and SOLAR--C.

\appendix

\begin{figure*}[h!]
\epsscale{1.15}
\plotone{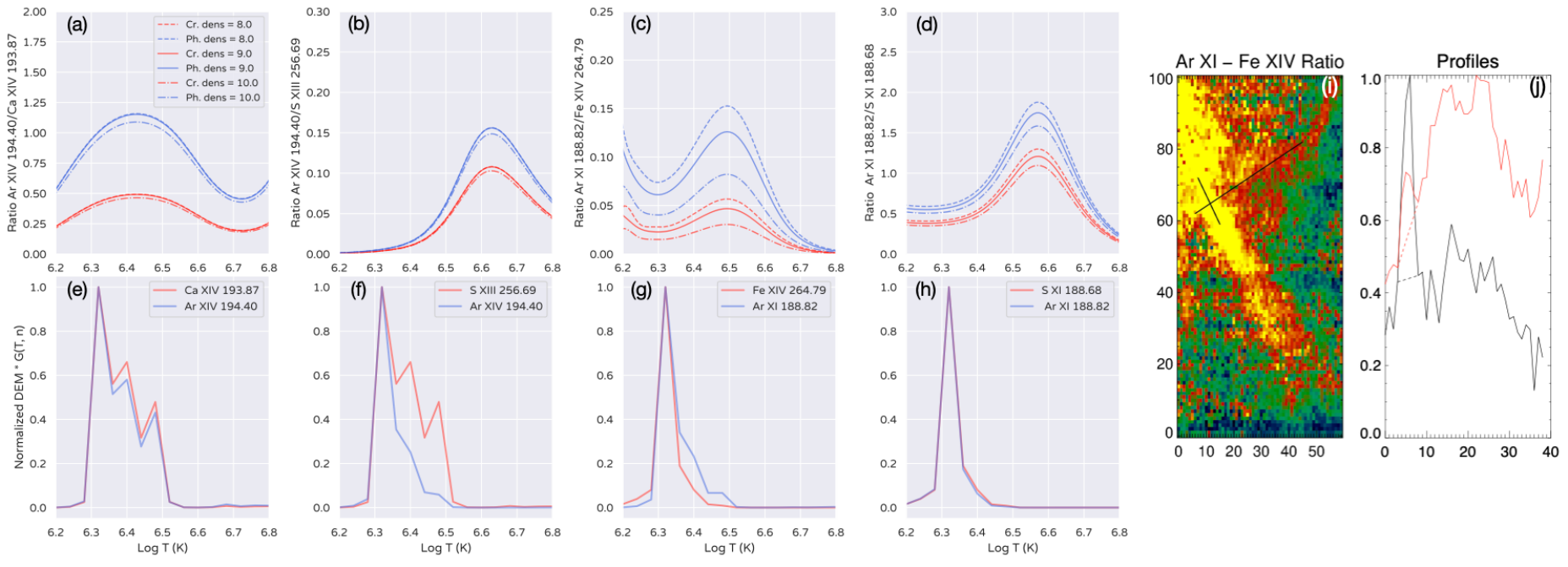}
\caption{Theoretical ratios for ion pairs (a) \ion{Ar}{14} 194.40 $\angstrom$--\ion{Ca}{14} 193.87 $\angstrom$, (b) \ion{Ar}{14} 194.40 $\angstrom$--\ion{S}{13} 256.69 $\angstrom$, (c) \ion{Ar}{11} 188.82 $\angstrom$--\ion{Fe}{14} 264.79 $\angstrom$, and (d) \ion{Ar}{11} 188.82 $\angstrom$--\ion{S}{11} 188.68 $\angstrom$ corresponding to the line intensity ratio maps shown in Figure \ref{f:eis}.  
Red lines in (a--d) are coronal and blue lines are photospheric ratios.
Normalized contribution function, G(T, n), convolved with DEM for I-FIP patches on June 14 at 00:14 UT (middle panels of Figure \ref{f:eis}) for (e) \ion{Ar}{14} 194.40 $\angstrom$ and \ion{Ca}{14} 193.87 $\angstrom$, (f) \ion{Ar}{14} 194.40 $\angstrom$ and \ion{S}{13} 256.69 $\angstrom$ (g) \ion{Ar}{11} 188.82 $\angstrom$ and \ion{Fe}{14} 264.79 $\angstrom$, (h) \ion{Ar}{11} 188.82 $\angstrom$ and \ion{S}{11} 188.68 $\angstrom$.
\ion{Ar}{11} 188.82 $\angstrom$--\ion{Fe}{14} 264.79 $\angstrom$ line intensity ratio map (i) and plot of \ion{Ar}{11} (black) and \ion{Fe}{14} (red) normalized intensity profiles (j) along the black line orthogonal to loops containing I-FIP plasma.
Note that the ratio map (i) is in pixels and is zoomed to the region surrounding the I-FIP plasma along the loops extending to the south-southwest.
\label{f:a_goft}} 
\end{figure*}

\section{Method to identify I-FIP plasma}

As discussed in Section \ref{s:eis_method}, line intensity ratios are used to identify I-FIP plasma in AR 11504.
The six emission lines listed in Table \ref{t:EISdetails} were fitted using the EISPAC fit templates: ar$\_$14$\_$194$\_$396.6c.template.h5, 
ca$\_$14$\_$193$\_$874.6c.template.h5,
ar$\_$11$\_$188$\_$806.3c.template.h5,
fe$\_$14$\_$264$\_$787.1c.template.h5,
s$\_$$\_$13$\_$256$\_$686.1c.template.h5, 
s$\_$$\_$11$\_$188$\_$675.3c.template.h5.
Line intensities were obtained from the fits to calculate the ratios in each pixel.
The intensity ratios were then compared to the theoretical ratios of the contribution functions at different densities over the temperature range of log T (MK) = [6.2, 6.8].
Theoretical ratios corresponding to the line intensity ratio maps in Figure \ref{f:eis} are plotted in Figure \ref{f:a_goft}(a--d). 
The contribution functions were calculated using the CHIANTI atomic database \citep{dere97} version 10 \citep{gdz21}.  Photospheric abundances are from \cite{scott15a,scott15b} and coronal abundances are from \cite{schmelz12}.  

In order to better understand the density and temperature effects for each ratio, we determined the density range and peak temperatures within the I-FIP patches on June 14 at 00:14 UT.
This was done for both diagnostics (2 MK and 4 MK).
The relevant spectra were averaged within each of the I-FIP patches before fitting. 
Log densities based on the \ion{Fe}{13} 202 $\angstrom$--\ion{Fe}{13} 203 $\angstrom$ density diagnostic line pair are [8, 9] cm$^{-3}$.
The corresponding peak temperatures were determined by convolving the contribution functions with the differential emission measure (DEM). 
See \cite{baker19} for details of the method.
The normalized contribution functions, G(T, n), convolved with DEMs are shown in Figure \ref{f:a_goft}(e-h).
The peak temperatures within the I-FIP patches are consistent for each line pair -- log T = 6.3.
Based on the measured densities and peak temperatures, safe levels for I-FIP is set at a line ratio $\geq$ 0.22 for \ion{Ar}{11} 188.82 $\angstrom$--\ion{Fe}{14} 264.79 $\angstrom$ and $\geq$ 1.75 for \ion{Ar}{14} 194.40 $\angstrom$--\ion{Ca}{14} 193.87 $\angstrom$, 50$\%$ above their respective maximum photospheric theoretical ratios.
A similar approach was taken for \ion{Ar}{11}-\ion{S}{11} and \ion{Ar}{14}--\ion{S}{13} line pairs.
I-FIP plasma is considered to be a ratio $\geq$2.82 for \ion{Ar}{11}-\ion{S}{11} and $\geq$0.23 for \ion{Ar}{14}--\ion{S}{13}.
These values do not account for a number of factors including: intensity measurement/calibration errors, changes to the contribution functions due to abundances, and the differences in the EIS instrument sensitivity across the long- and short-wave detectors \citep{gdz13,warren14}.

Further steps were taken to verify the I-FIP plasma observed along loops at 00:51 UT on June 15.
The ratios were calculated after background subtraction for each element to ensure that the signal-to-noise is sufficient.
Following the method of \cite{brooks13}, cross-loop intensity profiles were extracted along a number of lines perpendicular to the loop axis.
An example is shown in panel (j).
Intensities were then averaged for 5 stacked rows of pixels along each line.
The background-subtracted profile was fitted with a Gaussian function in order to obtain the integrated area under each curve.
This was done separately for the \ion{Ar}{11} and \ion{Fe}{14} intensities.
The ratio of the \ion{Ar}{11} and \ion{Fe}{14} integrated areas exceeded the I-FIP threshold of 0.22 in all cases,  thereby providing an extra level of confidence that the loops do indeed contain I-FIP plasma. 

\begin{acknowledgments}
Hinode is a Japanese mission developed and launched by ISAS/JAXA, collaborating with NAOJ as a domestic partner, and NASA and STFC (UK) as international partners. 
Scientific operation of Hinode is performed by the Hinode science team organized at ISAS/JAXA. 
This team mainly consists of scientists from institutes in the partner countries. 
Support for the post-launch operation is provided by JAXA and NAOJ (Japan), STFC (UK), NASA, ESA, and NSC (Norway). 
D.B. is funded under Solar Orbiter EUI Operations grant number ST/X002012/1 and Hinode Ops Continuation 2022-25 grant number ST/X002063/1.
L.v.D.G. acknowledges the Hungarian National Research, Development and Innovation Office grant OTKA K-131508.
A.W.J. acknowledges funding from the STFC Consolidated Grant ST/W001004/1. A.S.H.T. is supported by the European Space Agency (ESA) Research Fellowship. MM has been supported by the ASI-INAF agreement n. 2022-14-HH.0 and by the Italian agreement ASI-INAF 2021-12-HH.0 `Missione Solar-C EUVST-Supporto scientifico di Fase B/C/D'. 
DML is grateful to the Science Technology and Facilities Council for the award of an Ernest Rutherford Fellowship (ST/R003246/1).
The work of DHB was performed under contract to the Naval Research Laboratory and was funded by the NASA Hinode program.
JML was supported by the NASA Heliophysics Supporting Research Grant NNH22OB102 and by Basic Research Funds of the Office of Naval Research.
S.L.Y. is grateful to the Science Technology and Facilities Council for the award of an Ernest Rutherford
Fellowship (ST/X003787/1).
We recognise the collaborative and open nature of knowledge creation and dissemination, under the control of the academic community as expressed by Camille No\^{u}s at http://www.cogitamus.fr/indexen.html.
\end{acknowledgments}
\bibliographystyle{aasjournal}
\bibliography{refs}

\end{document}